\documentclass[aps,nofootinbib,commonaddress,preprint,showpacs]{revtex4}
\usepackage{amsmath}
\usepackage{amssymb}
\usepackage{graphicx}
\usepackage{color}

\begin{document}

\title{Surface plasmon polaritons in a semi-bounded degenerate plasma: role of spatial dispersion and collisions}
\author{Yu.~Tyshetskiy}
\email{y.tyshetskiy@physics.usyd.edu.au}
\affiliation{School of Physics, The University of Sydney, NSW 2006, Australia}
\author{R.~Kompaneets}
\affiliation{School of Physics, The University of Sydney, NSW 2006, Australia}
\author{S.V.~Vladimirov}
\affiliation{School of Physics, The University of Sydney, NSW 2006, Australia}
\affiliation{Metamaterials Laboratory, National Research University of Information Technology, Mechanics, and Optics, St Petersburg 199034, Russia}

\date{\today}

\begin{abstract}
Surface plasmon polaritons (SPPs) in a semi-bounded degenerate plasma (e.g., a metal) are studied using the quasiclassical mean-field kinetic model, taking into account the spatial dispersion of the plasma (due to quantum degeneracy of electrons) and electron-ion (electron-lattice, for metals) collisions. SPP dispersion and damping are obtained in both retarded ($\omega/k_z\sim c$) and non-retarded ($\omega/k_z\ll c$) regions, as well as in between. It is shown that the plasma spatial dispersion significantly affects the properties of SPPs, especially at short wavelengths (less than the collisionless skin depth, $\lambda\lesssim c/\omega_{pe}$). Namely, the collisionless (Landau) damping of SPPs (due to spatial dispersion) is comparable to the purely collisional (Ohmic) damping (due to electron-lattice collisions) in a wide range of SPP wavelengths, e.g., from $\lambda\sim20$~nm to $\lambda\sim0.8$~nm for SPP in gold at $T=293$~K, and from $\lambda\sim400$~nm to $\lambda\sim0.7$~nm for SPPs in gold at $T=100$~K. The spatial dispersion is also shown to affect, in a qualitative way, the dispersion of SPPs at short wavelengths $\lambda\lesssim c/\omega_{pe}$.
\end{abstract}
\maketitle


\newpage

\section{Introduction \label{sec:intro}}
As early as 1950s it has been demonstrated theoretically~\cite{Ritchie_1957}, and later confirmed experimentally~\cite{Powell_Swan_1959_1,Powell_Swan_1959_2,Otto_1968,Kretschmann_1968} that bounded metallic structures (e.g., thin films), as well as other bounded plasmas~\cite{Trivelpiece_Gould_1959,Guernsey_1969,Kondratenko_book}, can support a special type of collective oscillations of the plasma electrons, surface plasma waves, also called surface plasmons (SP), that propagate along the plasma boundaries and whose field and energy density are concentrated near the boundaries. 

Since then, there has been a significant advance in theoretical and experimental investigations of surface plasma waves and their applications in various bounded plasma structures, both in the field of plasma science (see Refs~\cite{Vladimirov_progress_1994,Denysenko_etal_2002,Shokri_2002} and references therein) and in the fields of condensed matter and surface science (see, e.g., Ref.~\cite{Pitarke_etal_2007} for a review). Currently, there is a renewed interest in surface plasmons due to their ability to concentrate light in subwavelength structures, enabling to create surface plasmon-based circuits that can couple photonics and electronics at nanoscale~\cite{Ozbay_2006, Barnes_etal_2003,Nomura_etal_2005}. This offers a route to faster and smaller devices, and even to new technologies employing surface plasmons~\cite{Plasmonics_Science_2010}. For example, one of the recent interesting advents in the new area of quantum nanoplasmonics is the development of the concept of spaser (a surface plasmon ``laser'')~\cite{SPASER_PRL_2003}, followed by its further development into a lasing spaser~\cite{Zheludev_etal_2008}, and by an experimental demonstration of a spaser-based nanolaser~\cite{Noginov_etal_2009,Lasers_go_nano}.

In view of these developments, understanding the properties of surface plasmons in various metallic (and semiconductor) structures, bounded by vacuum or dielectric, is thus important. This requires using models for the dynamic response of charge carriers in such structures to self-consistent electromagnetic fields, that appropriately take into account the relevant effects arising from quantum nature of the charge carriers in such structures and from their interaction with the underlying ion lattice. Such quantum effects may significantly alter the properties of the surface waves; see, e.g., Refs.~\cite{Vladimirov_Kohn_1994,Marklund_etal_new_quantum_limits}. 

In general, the electric field of a surface plasmon has both longitudinal and transverse components, i.e., a surface plasmon is, in general, an electromagnetic wave coupled with a collective oscillation of surface charge; such a hybrid electromagnetic surface wave is called a surface plasmon polariton (SPP). The properties (dispersion and damping) of SPP in a semi-bounded degenerate plasma (e.g., a metal) are a subject of this paper.

In the non-retarded limit, when the SPP phase velocity is much smaller than the speed of light, the SPP field becomes purely electrostatic. In this limit, the SPP corresponds to a longitudinal surface charge density wave; its properties are thus affected only by the longitudinal part $\varepsilon^l$ of the plasma dielectric response. Recently, the properties of such purely electrostatic surface plasmons (ESP) in a semi-bounded plasma with degenerate electrons were analyzed using the semi-classical kinetic model~\cite{Tysh_etal_ESSW_2012}. 
In particular, it was shown that the spatial dispersion of degenerate plasma has a significant effect on the frequency and damping rate of ESP at short wavelengths, $k_z\lambda_F\gtrsim 1$, where $k_z=2\pi/\lambda$ is the ESP wave vector component along the plasma boundary, $\lambda$ is the ESP wavelength, and $\lambda_F=v_F/\sqrt{3}\omega_{pe}$ is the Thomas-Fermi length ($v_F=\hbar\sqrt[3]{3\pi^2 n_e}/m_e$ is the Fermi velocity of plasma electrons, $\omega_{pe}=(4\pi e^2 n_e/m_e)^{1/2}$ is the electron plasma frequency, $-e$ and $m_e$ are the electron charge and mass, $\hbar$ is the reduced Planck constant, and $n_e$ is the electron number density in plasma; we use CGS units).

In this paper, we aim to generalize the work on electrostatic surface plasmons in a semi-bounded collisionless degenerate plasma~\cite{Tysh_etal_ESSW_2012}, and consider the properties of electromagnetic surface plasmons, or surface plasmon polaritons, in a semi-bounded degenerate plasma with electron-ion collisions (e.g., SPPs in a semi-bounded metal where the electron-lattice collisions play a significant role). In particular, we consider effects of plasma spatial dispersion, and collisions, on SPP dispersion and damping. We show that the collisionless damping of SPPs (due to spatial dispersion of a metal) is comparable, or even exceeds the purely collisional (Ohmic) damping of SPPs (due to electron-lattice collisions) in a wide range of wavelengths, e.g., from $\lambda\sim 20$~nm to $\lambda\sim 0.8$~nm for SPPs in gold at $T=293$~K, and from $\lambda\sim400$~nm to $\lambda\sim0.7$~nm for SPPs in gold at $T=100$~K. The spatial dispersion is also shown to affect the SPP spectrum, especially at short wavelengths (less than the collisionless skin depth, $\lambda\lesssim c/\omega_{pe}$), where the spatial dispersion changes SPP spectrum in a qualitative way. We conclude that the spatial dispersion (due to quantum degeneracy of plasma electrons) is rather important for damping, and, to a lesser extent, for dispersion of SPPs in a semi-bounded degenerate plasma (e.g., a bounded metal), and should be taken into account in the relevant SPP models.

\section{Method}
\subsection{Model and Assumptions \label{sec:model}}
We consider a semi-bounded, nonrelativistic plasma (e.g., a metal) with degenerate mobile electrons ($T_e\ll\epsilon_F$, where $T_e$ is the electron temperature in energy units, $\epsilon_F=\hbar^2(3\pi^2n_e)^{2/3}/2 m_e$ is the electron Fermi energy), and immobile ions; the equilibrium number densities of electrons and ions are equal, $n_{0e}=n_{0i}=n_0$ (quasineutrality). The plasma is assumed to be confined to a region $x<0$, with mirror reflection of plasma particles at the boundary $x=0$ separating the plasma from a vacuum at $x>0$. In the absence of fields, the equilibrium distribution function of plasma electrons $f_{0e}(\mathbf{p})$ is defined by an isotropic Fermi-Dirac distribution, which in the limit $T_e\ll\epsilon_F$ reduces to
\begin{equation}
f_{0e}(p)=\frac{2}{(2\pi\hbar)^3}\left\{1 + \exp\left[\frac{p^2/2m_e - \epsilon_F(n_e)}{T_e}\right] \right\}^{-1} = \frac{2}{(2\pi\hbar)^3} \sigma\left[p_F(n_e)-p\right],  \label{eq:f0e}
\end{equation}
where $p_F(n_e)=\sqrt{2m_e\epsilon_F(n_e)}$ is the electron Fermi momentum, $\sigma(x)$ is the Heaviside step function.

Following the discussion of Ref.~\cite{Tysh_etal_ESSW_2012}, we adopt here the quasiclassical kinetic description of plasma electrons in terms of the 1-particle distribution function $f_e(\mathbf{r,p},t)$~\cite{Vlad_Tysh_UFN_2011}, whose evolution is described by the kinetic equation
\begin{equation}
\frac{\partial f_e}{\partial t} + \frac{\mathbf{p}}{m_e}\cdot\frac{\partial f_e}{\partial\mathbf{r}} - e\left(\mathbf{E} + \frac{\mathbf{v\times B}}{c}\right)\cdot\frac{\partial f_e}{\partial\mathbf{p}} = \sum_{\alpha}{I_{e\alpha}(f_e,f_\alpha)},  \label{eq:kinetic}
\end{equation}
where $I_{e\alpha}(f_e,f_\alpha)$ describes collisions of electrons with particles of sort $\alpha=e,i$, i.e., with electrons and ions. Here we are only interested in the electron-ion collisions (due to coherent scattering of electrons on the ions, which in metals is equivalent to electron-phonon scattering), which transfer the energy of quiver motion of electrons in the field of collective plasma oscillations (in particular, in the field of SPP) to the ``ionic thermal bath'' (e.g., ion lattice in metals), resulting in SPP damping.

\subsection{Electron-ion collisions}
In general, collisions between degenerate electrons and non-degenerate classical ions can be described by the quantum Lennard-Balescu collision integral~\cite{Wyld_Pines_1962,Kremp_book}
\begin{eqnarray}
I_{ei}(\mathbf{p}) = \int\frac{d\mathbf{p}^\prime d\mathbf{p}_id\mathbf{p}_i^\prime}{(2\pi\hbar)^6}\frac{2\pi}{\hbar}\left|V_{ei}(\mathbf{p-p'})\right|^2\delta\left[\epsilon_e(\mathbf{p}^\prime) + \epsilon_i(\mathbf{p}_i^\prime) - \epsilon_e(\mathbf{p}) - \epsilon_i(\mathbf{p}_i)\right] \nonumber \\
\times\delta\left[\mathbf{p}+\mathbf{p}_i-\mathbf{p}^\prime-\mathbf{p}_i^\prime\right]\left\{f_e(\mathbf{p}^\prime)\left[1-f_e(\mathbf{p})\right]f_i(\mathbf{p}_i^\prime) - f_e(\mathbf{p})\left[1-f_e(\mathbf{p}^\prime)\right]f_i(\mathbf{p}_i) \right\}, \label{eq:I_ei_Balescu}
\end{eqnarray}
where $\epsilon_{e,i}(\mathbf{p})=|\mathbf{p}|^2/2m_{e,i}$, and $V_{ei}(\mathbf{q})$ is the Fourier transform of the dynamically screened electron-ion interaction potential, 
\[ 
V_{ei}(\mathbf{q}) = \int d\mathbf{r}\exp\left(-\frac{i}{\hbar}\mathbf{q\cdot r}\right) V_{ei}(\mathbf{r}) = -\frac{4\pi\hbar^2e^2 Z_i}{q^2 \varepsilon^l(\omega_q,\mathbf{q}/\hbar)},
\]
where $\varepsilon^l(\omega,\mathbf{k})$ is the longitudinal dielectric response of the plasma, $\omega_q=\hbar^{-1}\left[\epsilon_e(\mathbf{p+q})-\epsilon_e(\mathbf{p})\right]$, $\mathbf{q}$ is the momentum transfered in the collision, and $Z_i$ is the ionization number of plasma ions.

In a weak spatially uniform oscillating field, $\mathbf{E}=\mathbf{E}_0\exp(-i\omega t)$, the electron distribution function can be approximated as $f_e(\mathbf{r,p},t) = f_{0e}(p) + \delta f_e(\mathbf{r,p},t)$, with a perturbation $\delta f_e(\mathbf{r,p},t) = g(\mathbf{r},p)\cos\theta$, where $\theta$ is the angle between $\mathbf{p}$ and $\mathbf{E}$, and where $|g(\mathbf{r},p)|\ll f_{0e}(p)$ is assumed (due to the field being weak). In this case the electron-ion collision integral (\ref{eq:I_ei_Balescu}) reduces to~\cite{Kremp_book}
\begin{equation}
I_{ei}(\mathbf{p}) = -\frac{\delta f_e(\mathbf{r,p},t)}{\tau_{ei}(\mathbf{p})} = -\frac{ f_e - f_{0e}}{\tau_{ei}(\mathbf{p})},  \label{eq:I_eq_reduced}
\end{equation}
where $\tau_{ei}^{-1}(\mathbf{p})$ is the electron-ion collision frequency for electrons with momentum $\mathbf{p}$, defined as~\cite{Kremp_book, Ziman_1961}
\begin{equation}
\tau_{ei}^{-1}(\mathbf{p}) \approx \frac{n_{0i}m_e}{2p^3}\int_0^{2p/\hbar}\frac{dk}{2\pi}a(k) k^3 \left|\frac{4\pi Z_i e^2}{k^2\varepsilon^l(0,k)}\right|^2.  \label{eq:tau_ei_gen}
\end{equation}
Note that (\ref{eq:tau_ei_gen}) is obtained by assuming static screening, $\varepsilon(\omega,k)\approx\varepsilon(0,k)$, for simplicity. The structure function $a(k)$ in (\ref{eq:tau_ei_gen}) depends on the arrangement of ions in the system considered, and can be obtained from experimental measurements of x-ray scattering on the corresponding medium~\cite{Alekseev_UFN}.

Yet in case of a weak \textit{non-uniform} oscillating field (e.g., the self-consistent field of plasma collective oscillations, such as SPPs), the perturbation $\delta f_e(\mathbf{r,p},t)$ is no longer proportional to $\cos\theta$, and thus the integral (\ref{eq:I_ei_Balescu}) no longer reduces to the simple form (\ref{eq:I_eq_reduced}). Nevertheless, the form (\ref{eq:I_eq_reduced}) is rather attractive due to its simplicity, and it is thus desirable to use this form in our model, especially since it becomes \textit{exact} for fields that are close to uniform, i.e., for plasma perturbations with $k\lambda_F\ll 1$. 
We therefore adopt a \textit{model} collision integral describing collisions of electrons with the heavy plasma ions (e.g., with the ion lattice in metals) in the form of Bhatnagar-Gross-Krook (BGK) integral~\cite{ABR_book}:
\begin{equation}
I_{ei}^{BGK}(\mathbf{p}) = -\nu_{ei}\left[f_e - f_{0e}(n_e,T_e)\right],  \label{eq:I_ei_BGK}
\end{equation}
where $f_{0e}(n_e,T_e)$ is the quasi-equilibrium distribution of electrons given by (\ref{eq:f0e}) with $\epsilon_F(n_e)=\hbar^2(3\pi^2n_e)^{2/3}/2m_e$, in which $n_e$ and $T_e$ are the \textit{perturbed} electron density and temperature, defined as
\begin{eqnarray}
n_e(\mathbf{r},t)&=&\int f_e(\mathbf{r,p},t) d^3\mathbf{p}, \\
T_e(\mathbf{r},t)&=&\frac{m_e}{2 n_e(\mathbf{r},t)}\int v^2 f_e(\mathbf{r,p},t) d^3\mathbf{p}.
\end{eqnarray}
The model collision integral (\ref{eq:I_ei_BGK}) conserves the number of particles, and energy, and mimics the relaxation of electron distribution towards the quasi-equilibrium distribution $f_{0e}(n_e,T_e)$, occurring over the characteristic time $\nu_{ei}^{-1}$. The value of $\nu_{ei}$ can be estimated as $\nu_{ei}\sim1/\tau_{ei}(p_F)$, with $\tau_{ei}(p_F)$ from (\ref{eq:tau_ei_gen}), for which the knowledge of the structure function $a(k)$ is required. However, for a crude estimate (sufficient for our model BGK integral), we can simply use the dc resistivity of the bulk plasma, $\rho=\nu_{ei} m_e/e^2 n_{0e}$, which is available for most metals from experimental measurements. This gives
\begin{equation}
\frac{\nu_{ei}}{\omega_{pe}} = \varepsilon_0 \omega_{pe} \rho\ \text{ (in SI units).}  \label{eq:nu_ei}
\end{equation}
For example, for gold at room temperature ($T=293$~K) $\rho\approx 2.2\times10^{-8}$ Ohm$\cdot$m~\cite{Phys_Chem_book}, $\omega_{pe}\approx 1.3\times10^{16}$~s$^{-1}$~\cite{Phys_Chem_book,Ordal_etal_1985}, which gives
\begin{equation}
\left.\frac{\nu_{ei}}{\omega_{pe}}\right|_{\text{Au,{\it T}=293 K}} \approx 2.5\times10^{-3}.  \label{eq:nu_gold}
\end{equation}

\subsection{Dispersion equation for SPP}
Introducing a small perturbation $\delta f_e = f_e-f_{0e}$ associated with a weak electromagnetic field $\mathbf{E, B}$, and linearizing the kinetic equation (\ref{eq:kinetic}) with $I_{ee}=0$ and $I_{ei}$ defined by (\ref{eq:I_ei_BGK}), we obtain the equation for $\delta f_e$ in the form
\begin{equation}
\left(\frac{\partial}{\partial t} + \frac{\mathbf{p}}{m_e}\cdot\frac{\partial}{\partial\mathbf{r}}\right)\delta f_e - e\mathbf{E}\cdot\frac{\partial f_{0e}}{\partial\mathbf{p}} = -\nu_{ei}\left[\delta f_e + \frac{2}{3}\frac{\epsilon_F(n_0)}{n_0}\left.\frac{\partial f_{0e}}{\partial\epsilon}\right|_{n_0}\int\delta f_e d^3\mathbf{p}^\prime\right],  \label{eq:kin_lin}
\end{equation}
where we have used the isothermal approximation for the BGK collision integral (see Appendix~\ref{sec:BGK_isothermal}). Eq.~(\ref{eq:kin_lin}) is complemented with the Maxwell's equations, with charge and current densities defined in terms of $\delta f_e$ as $\rho_q=-e\int{\delta f_e d^3\mathbf{p}}$, $\mathbf{j}=-(e/m_e)\int{\mathbf{p} \delta f_e d^3\mathbf{p}}$, with imposed continuity of $\mathbf{E, B}$ components parallel to the plasma boundary $x=0$, across the plasma-vacuum interface. Assuming a specular reflection of plasma particles (electrons in our case) at $x=0$, and following the standard procedure (see, e.g., Ref.~\cite{ABR_book}), we obtain the dispersion relation for TM-polarized SPP (note that TE-polarized SPPs are not allowed in the considered system):
\begin{eqnarray}
\sqrt{k_z^2-\frac{\omega^2}{c^2}} + \frac{1}{\pi}\int_{-\infty}^{+\infty}\frac{dk_x}{k^2}\left\{\frac{k_z^2}{\varepsilon^l(\omega,\mathbf{k})} - \frac{\omega^2 k_x^2}{c^2k^2-\omega^2\varepsilon^{tr}(\omega,\mathbf{k})} \right\} = 0,  \label{eq:SPP_dispersion}
\end{eqnarray}
where $\omega$ and $\mathbf{k}$ are the frequency and wave vector of TM SPP, respectively, $k_x$ and $k_z$ are the components of $\mathbf{k}$ perpendicular and parallel to the boundary $x=0$, respectively, so that $k\equiv|\mathbf{k}|=\sqrt{k_x^2+k_z^2}$. Note that in the non-retarded limit, $\omega/k\ll c$, the dispersion equation (\ref{eq:SPP_dispersion}) for SPP reduces to the dispersion equation for ESW, studied in Ref.~\cite{Tysh_etal_ESSW_2012}:
\[
1 + \frac{k_z}{\pi}\int_{-\infty}^{+\infty}\frac{dk_x}{k^2 \varepsilon^l(\omega,k)} = 0.
\]

The plasma properties with respect to the weak electromagnetic field of SPP are entirely contained in the longitudinal and transverse linear dielectric response functions $\varepsilon^l(\omega,\mathbf{k})$ and $\varepsilon^{tr}(\omega,\mathbf{k})$, which, for the model (\ref{eq:kin_lin}) with isotropic equilibrium distribution (\ref{eq:f0e}), are found to be~\cite{ABR_book}
\begin{eqnarray}
\varepsilon^l(\omega,k) &=& 1 + 3\frac{\omega_{pe}^2}{k^2v_F^2}\left[1-\frac{\omega+i\nu_{ei}}{2kv_F}\ln\left(\frac{\omega+i\nu_{ei}+kv_F}{\omega+i\nu_{ei}-kv_F}\right)\right] \nonumber \\
&\times& \left[1-\frac{i\nu_{ei}}{2kv_F}\ln\left(\frac{\omega+i\nu_{ei}+kv_F}{\omega+i\nu_{ei}-kv_F}\right)\right]^{-1}, \label{eq:eps_l}\\
\varepsilon^{tr}(\omega,k) &=& 1 - \frac{3}{2}\frac{\omega_{pe}^2}{\omega\left(\omega+i\nu_{ei}\right)}\left\{1+\left[
\frac{\left(\omega+i\nu_{ei}\right)^2}{k^2v_F^2} - 1\right]\right. \nonumber \\
&\times&\left.\left[1-\frac{\omega+i\nu_{ei}}{2kv_F}\ln\left(\frac{\omega+i\nu_{ei}+kv_F}{\omega+i\nu_{ei}-kv_F}\right)\right] \right\}. \label{eq:eps_tr}
\end{eqnarray}

\section{Properties of surface plasmon polaritons}
Eq.~(\ref{eq:SPP_dispersion}) with $\varepsilon^l(\omega,\mathbf{k})$ and $\varepsilon^{tr}(\omega,\mathbf{k})$ defined by Eqs~(\ref{eq:eps_l})-(\ref{eq:eps_tr}) describes dispersion and damping of SPP in a semi-bounded degenerate plasma (e.g., a metal) with electron-ion collisions approximated by the BGK integral in the right-hand side of (\ref{eq:kin_lin}). Below we analyze the effects of plasma spatial dispersion and electron-ion collisions on SPP properties.

To see how the spatial dispersion ($k$-dependence of plasma responses $\varepsilon^l(\omega,\mathbf{k})$ and $\varepsilon^{tr}(\omega,\mathbf{k})$, due to electron velocity spread $v_F$ arising from Pauli blocking) and electron-ion collisions affect the dispersion and damping of SPPs in the considered system, it is instructive to compare the following three models:
\begin{enumerate}
\item ``Local collisional'' approximation for the medium, in which the electron-ion collisions are retained, while the spatial dispersion is neglected, by formally taking the limit $v_F\to 0$ in (\ref{eq:eps_l})-(\ref{eq:eps_tr}), which gives
\begin{equation}
\varepsilon^l(\omega,k)=\varepsilon^{tr}(\omega,k) = 1 - \frac{\omega_{pe}^2}{\omega\left(\omega+i\nu_{ei}\right)}.
\end{equation}
This model describes purely collisional effects on SPP dispersion and damping. The SPP frequency and damping rate obtained in the ``local collisional'' approximation are denoted by $\omega_{v_F=0, \nu_{ei}>0}$ and $\gamma_{v_F=0, \nu_{ei}>0}$, respectively.
\item ``Nonlocal collisionless'' approximation, in which the spatial dispersion is retained, while the electron-ion collisions are neglected by taking the limit $\nu_{ei}\to 0$ in (\ref{eq:eps_l})-(\ref{eq:eps_tr}). This model describes the effects of spatial dispersion on SPP properties, e.g., collisionless damping of SPPs. The SPP frequency and damping rate obtained in the ``nonlocal collisionless'' approximation are denoted by $\omega_{v_F>0, \nu_{ei}=0}$ and $\gamma_{v_F>0, \nu_{ei}=0}$, respectively.
\item ``Nonlocal collisional'', i.e., the full model, in which both spatial dispersion and collisions are retained in (\ref{eq:eps_l})-(\ref{eq:eps_tr}). This model describes the combined effects of spatial dispersion and collisions on SPP properties. The SPP frequency and damping rate obtained in the ``nonlocal collisional'' approximation are denoted by $\omega_{v_F>0, \nu_{ei}>0}$ and $\gamma_{v_F>0, \nu_{ei}>0}$, respectively.
\end{enumerate}

Only the weakly damped SPPs are of physical interest. Their dispersion and damping are obtained in the same way as described in detail in Ref.~\cite{Tysh_etal_ESSW_2012}.
\subsection{Dispersion}
\begin{figure}
\includegraphics[width=4.0in]{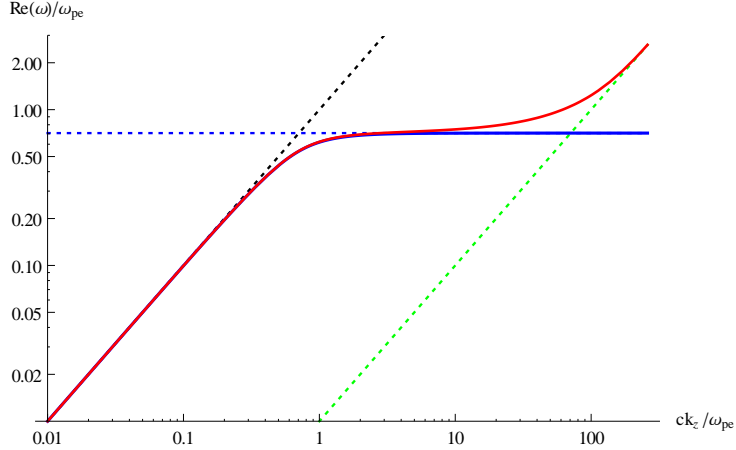}
\caption{\label{fig:dispersion} (Color online) SPP spectra for gold at room temperature ($T=293$~K), obtained from the ``local collisional'' model (solid blue curve), ``nonlocal collisional'' (full) model (solid red curve), and ``nonlocal collisionless'' model (not seen due to exact matching with the red curve). The light dispersion $\omega=ck_z$ is shown with the dotted black line, the ``cold plasma limit'' $\omega=\omega_{pe}/\sqrt{2}$ is shown with the dotted blue line, and the zero sound asymptote (\ref{eq:omega_wc_kz>>1}) is shown with the dotted green line. 
The gold parameters are~\cite{Phys_Chem_book}: $\omega_{pe}=1.3\times10^{16}$~s$^{-1}$, $\nu_{ei}/\omega_{pe}=2.5\times10^{-3}$ [see Eq.~(\ref{eq:nu_gold})], $v_F/c=5\times10^{-3}$.}
\end{figure}

The spectra of SPP $\omega(k_z)$ obtained from the three models introduced above (``local collisional'', ``nonlocal collisionless'', and ``nonlocal collisional'') are shown in Fig.~\ref{fig:dispersion}, for the case of gold at room temperature. 
It is seen that the SPP phase velocities given by all three models are smaller than the speed of light, $\omega/k_z<c$; i.e., the surface plasmons in a semi-bounded plasma with an ideal boundary are essentially non-radiative (they cannot decay by emitting a photon, and cannot be excited by a photon incident on the plasma surface). The dispersion curves for the ``nonlocal collisionless'' and ``nonlocal collisional'' models match, hence the collisions do not affect the SPP spectrum. The effect of the spatial dispersion on SPP spectrum, however, is significant, especially at short SPP wavelengths ($\lambda\lesssim 15$~nm for gold at room temperature), as seen from Fig.~\ref{fig:dispersion}. The asymptotic expressions for SPP frequencies $\omega(k_z)$, given by the ``local collisional'' and ``nonlocal collisional'' models match:
\[
\omega_{v_F=0, \nu_{ei}>0} = \omega_{v_F>0, \nu_{ei}>0} = ck_z - O\left[(ck_z)^2\right],\ \ \ ck_z\ll\omega_{pe},
\]
i.e., the spatial dispersion does not affect the SPP spectrum significantly at large wavelengths. However, at short wavelengths, the ``local collisional'' and ``nonlocal collisional'' models yield qualitatively different asymptotes for SPP frequencies: without the spatial dispersion, the SPP frequency approaches the ``cold plasma limit'' \begin{eqnarray}
\omega_{v_F=0, \nu_{ei}>0} &=& \frac{\omega_{pe}}{\sqrt{2}},\ \ \ ck_z\gg\omega_{pe}, \label{eq:omega_cc_kz>>1}
\end{eqnarray}
while accounting for the spatial dispersion leads to a qualitatively different SPP frequency asymptote~\cite{Tysh_etal_ESSW_2012}:
\begin{eqnarray}
\omega_{v_F>0, \nu_{ei}>0} &=& k_z v_F\left(1+2\exp\left[-2-\frac{4}{3}\frac{k_z^2v_F^2}{\omega_{pe}^2}\right]\right)\approx k_zv_F,\ \ \ ck_z\gg\omega_{pe}.  \label{eq:omega_wc_kz>>1}
\end{eqnarray}
It is interesting to note that this part of the SPP spectrum (\ref{eq:omega_wc_kz>>1}) is exponentially close to the spectrum of volume zero sound mode, propagating along the boundary (with $k=k_z$) in an \textit{uncharged} Fermi gas.

\subsection{Damping}
\begin{figure}
\includegraphics[width=4.0in]{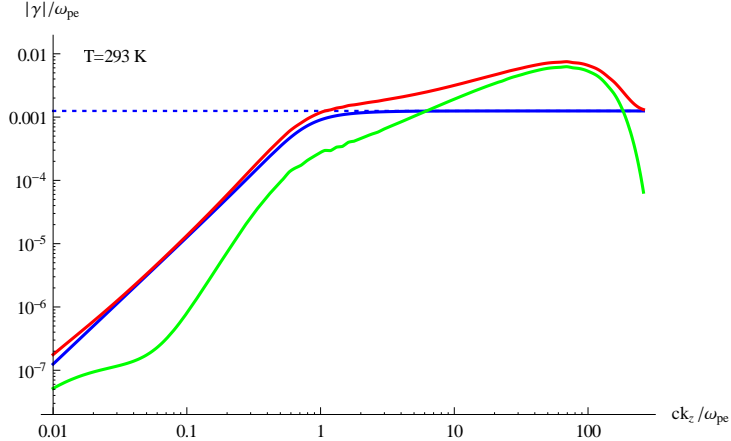}
\caption{\label{fig:damping_gold_293K} (Color online) SPP damping for gold at room temperature ($T=293$~K), obtained from the ``local collisional'' model (solid blue curve), ``nonlocal collisionless'' model (solid green curve), and ``nonlocal collisional'' (full) model (solid red curve). The dotted blue line marks the maximum purely collisional damping $\gamma=-\nu_{ei}/2$ [see Eq.~(\ref{eq:max_coll_damping})]. 
The parameters are the same as in Fig.~\ref{fig:dispersion}.}
\end{figure}

The SPP damping rates $\gamma(k_z)$, obtained from the three models (``local collisional'', ``nonlocal collisionless'', and ''nonlocal collisional'') are shown in Fig.~\ref{fig:damping_gold_293K}, for the case of gold at room temperature. 
The purely collisional (ohmic) SPP damping rate, given by the ``local collisional'' model, has the following asymptotes:
\begin{eqnarray}
\gamma_{v_F=0, \nu_{ei}>0} &=& -\frac{\nu_{ei}}{2}\frac{c^2 k_z^2}{\omega_{pe}^2},\ \ \ ck_z\ll\omega_{pe}, \\
\gamma_{v_F=0, \nu_{ei}>0} &=& -\frac{\nu_{ei}}{2},\ \ \ \ \ ck_z\gg\omega_{pe}.  \label{eq:max_coll_damping}
\end{eqnarray}
The spatial dispersion of the plasma (due to quantum degeneracy of electrons) leads to collisionless SPP damping, which exhibits a non-monotonic dependence on $k_z$, as seen in Fig.~\ref{fig:damping_gold_293K}: at small $k_z$, the collisionless damping rate monotonically increases with $k_z$, reaching a distinct maximum of $\gamma_{v_F>0, \nu_{ei}=0}\approx 6.2\times10^{-3}\ \omega_{pe}$ at $c k_z/\omega_{pe}\approx 0.7 c/v_F$, and then \textit{decreases} monotonically with $k_z$ for $c k_z/\omega_{pe}>0.7c/v_F$, quickly approaching zero at $ck_z\gg\omega_{pe}$. 

It might seem somewhat counter-intuitive that SPPs are subject to collisionless damping at all, despite the fact that their phase velocity is larger than the maximum electron velocity in degenerate plasma, $\omega/k_z>v_F$. Indeed, it is well known that the \textit{volume} plasmons in degenerate plasma, described within the quasiclassical model employed here (with quantum recoil ignored), are not subject to collisionless damping at all, exactly due to their phase velocity being larger than the maximum electron velocity in degenerate plasma, $\omega/k>v_F$, so that no plasma electrons can be in resonance with the plasmons. However, a surface plasmon is in fact a result of interference of many (a continuum, in case of SPP in a semi-bounded plasma) ``virtual'' volume plasmons with a given $k_z$, but with all possible $k_x$ ranging from minus to plus infinity [see Eq.~(\ref{eq:SPP_dispersion})]. As a result, the surface plasmon consists of virtual volume plasmons with phase velocities that are both larger and smaller than $v_F$, and the latter are subject to resonant collisionless damping on plasma electrons, resulting in a finite collisionless damping of the resultant surface plasmon, even despite its phase velocity $\omega/k_z$ exceeding the electron Fermi velocity.

\begin{figure}
\includegraphics[width=4.0in]{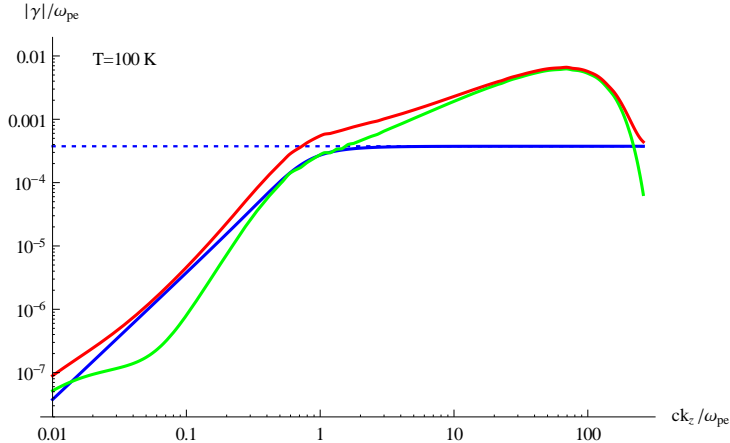}
\caption{\label{fig:fig:damping_gold_100K} (Color online) SPP damping for gold at $T=100$~K ($\rho=6.5\times10^{-9}$~Ohm m~\cite{Phys_Chem_book}, corresponding to $\nu_{ei}/\omega_{pe}\sim 7.5\times10^{-4}$), obtained from the ``local collisional'' model (solid blue curve), ``nonlocal collisionless'' model (solid green curve), and ``nonlocal collisional'' (full) model (solid red curve). The dotted blue line marks the maximum purely collisional damping $\gamma=-\nu_{ei}/2$ [see Eq.~(\ref{eq:max_coll_damping})].}
\end{figure}

As seen from Fig.~\ref{fig:damping_gold_293K}, for gold at room temperature ($T=293$~K), as well as for other metals with similar properties (e.g., silver, aluminium), the collisionless damping rate is equal to, or larger than the purely collisional damping rate, for SPP wavelengths ranging (for the case of gold) from extreme ultraviolet ($ck_z/\omega_{pe}\sim 6$, corresponding to $\lambda\sim 20$~nm) to soft x-ray ($ck_z/\omega_{pe}\sim 2\times10^2$, corresponding to $\lambda\sim 0.8$~nm). The full SPP damping rate $\gamma_{v_F>0, \nu_{ei}>0}$, due to the combined effect of collisions and spatial dispersion, given by the full ``nonlocal collisional'' model, is approximately (for $\nu_{ei}/\omega_{pe}\ll1$, which is the case for metals) a sum of the purely collisional $\gamma_{v_F=0, \nu_{ei}>0}$ and the purely collisionless $\gamma_{v_F>0, \nu_{ei}>0}$ damping rates, and is dominated by the collisionless damping at short SPP wavelengths, as seen from Fig.~\ref{fig:damping_gold_293K}. At lower metal temperatures, the resistivity and, correspondingly, $\nu_{ei}$ drop~\cite{Phys_Chem_book}, resulting in widening of the wavelength range at which the SPP collisionless damping dominates over the collisional damping. For example, at $T=100$~K, we have $\rho=0.65\times10^{-8}$~Ohm m~\cite{Phys_Chem_book}, corresponding to $\nu_{ei}/\omega_{pe}\sim7.5\times10^{-4}$ (see Eq.~(\ref{eq:nu_ei})), and the range of SPP wavelengths at which the collisionless damping dominates extends all the way from $\lambda\sim 400$~nm to $\lambda\sim0.7$~nm, as seen in Fig.~\ref{fig:fig:damping_gold_100K}.

\section{Conclusion}
In this paper, surface plasmon polaritons in a semi-bounded degenerate plasma (e.g., a metal) were studied, using the quasiclassical kinetic model with the BGK collision integral modeling electron-ion (electron-lattice for metals) collisions. Spectrum and damping rate are obtained for SPP in both retarded and non-retarded cases, thus extending the results of Ref.~\cite{Tysh_etal_ESSW_2012} in which only the non-retarded limit was treated. The effects of plasma spatial dispersion (due to velocity spread of degenerate electrons) and of electron-lattice collisions on SPP dispersion and damping are studied. It is found that the collisions do not affect SPP spectrum, but contribute to SPP damping. The spatial dispersion, however, significantly affects both spectrum and damping of SPPs, especially at short wavelengths. Namely, it is due to the plasma spatial dispersion that the SPP spectrum changes from the ``cold plasma limit'' $\omega=\omega_{pe}/\sqrt{2}$ to the ``zero sound'' asymptote $\omega\approx k_z v_F$ at short wavelengths, $\lambda\lesssim c/\omega_{pe}$. In metals, the damping of SPP is approximately a sum of purely collisional (Ohmic) and collisionless (Landau) damping, and is dominated by the spatial dispersion at a wide range of wavelengths, especially at low metal temperatures. For example, in gold at $T=293$~K, the total SPP damping is dominated by the collisionless damping (due to spatial dispersion arising from quantum degeneracy of conduction electrons) for SPP wavelengths ranging from $\lambda\sim20$~nm to $\lambda\sim0.8$~nm, while in gold at $T=100$~K (e.g., cooled in liquid nitrogen) the total SPP damping is dominated by the collisionless damping in a much wider range of SPP wavelengths, spanning from $\lambda\sim 400$~nm to $\lambda\sim0.7$~nm. We thus conclude that the spatial dispersion is important for SPP properties, especially in metals at low temperatures, and should be taken into account in the relevant models.

\acknowledgments{This work was supported by the Australian Research Council. R.K. acknowledges the receipt of a Professor Harry Messel Research Fellowship funded by the Science Foundation for Physics within the University of Sydney.}

\appendix
\section{BGK integral in isothermal approximation \label{sec:BGK_isothermal}}
The BGK collision integral is defined by (\ref{eq:I_ei_BGK}) with $f_{0e}(n_e,T_e)$ being the quasi-equilibrium Fermi-Dirac distribution depending on the \textit{perturbed} electron density $n_e$ and temperature $T_e$, which for $T_e\ll\epsilon_F(n_e)$ is given by (\ref{eq:f0e}). At equilibrium we have $f_e=f_{0e}(n_{0e},T_{0e})$, where $n_{0e}$ and $T_{0e}$ are the equilibrium electron density and temperature. Introducing a small perturbation of the equilibrium distribution $f_e=f_{0e}+\delta f_e$, $|\delta f_e|\ll f_{0e}$, we also introduce small perturbations of the electron density and temperature, $n_e=n_{0e}+\delta n_e$ ($|\delta n_e|\ll n_{0e}$), and $T_e=T_{0e}+\delta T_e$ ($|\delta T_e|\ll T_{0e}$). In the isothermal approximation, we neglect the temperature perturbations, assuming $T_e=T_{0e}$ (with $T_{0e}\ll\epsilon_F(n_e)$). Then, expanding $f_{0e}(n_e,T_{0e})$ on $\delta n_e$ and linearizing with respect to the small perturbation, we obtain
\begin{eqnarray}
I_{ei}^{BGK} = -\nu_{ei}\left[\delta f_e - \left.\frac{\partial f_{0e}(n_e,T_{0e})}{\partial n_e}\right|_{n_{0e},T_{0e}} \delta n_e\right],
\end{eqnarray}
with $f_{0e}(n_e,T_{0e})$ defined by (\ref{eq:f0e}). Differentiating, we obtain
\begin{eqnarray}
I_{ei}^{BGK} = -\nu_{ei}\left\{\delta f_e - \frac{2}{3}\frac{\delta n_e}{n_{0e}}\frac{\epsilon_{F0}}{T_{0e}} f_{0e}(n_{0e},T_{0e})\left[1-\frac{(2\pi\hbar)^3}{2}f_{0e}(n_{0e},T_{0e})\right]\right\},
\end{eqnarray} 
where $\epsilon_{F0}=\epsilon_{F}(n_{0e})$. Finally, noting that
\[
\frac{f_{0e}}{T_{0e}}\left[1-\frac{(2\pi\hbar)^3}{2}f_{0e}\right] = -\frac{\partial f_{0e}}{\partial\epsilon},
\]
where $\epsilon=p^2/2m_e$ is the electron kinetic energy, we obtain
\begin{eqnarray}
I_{ei}^{BGK} = -\nu_{ei}\left[\delta f_e + \frac{2}{3}\epsilon_{F0}\left.\frac{\partial f_{0e}}{\partial\epsilon}\right|_{n_{0e},T_{0e}}\frac{\delta n_e}{n_{0e}}\right],
\end{eqnarray} 
with $\delta n_e = \int{\delta f_e d^3\mathbf{p}}$.


\begin{thebibliography}{31}
\expandafter\ifx\csname natexlab\endcsname\relax\def\natexlab#1{#1}\fi
\expandafter\ifx\csname bibnamefont\endcsname\relax
  \def\bibnamefont#1{#1}\fi
\expandafter\ifx\csname bibfnamefont\endcsname\relax
  \def\bibfnamefont#1{#1}\fi
\expandafter\ifx\csname citenamefont\endcsname\relax
  \def\citenamefont#1{#1}\fi
\expandafter\ifx\csname url\endcsname\relax
  \def\url#1{\texttt{#1}}\fi
\expandafter\ifx\csname urlprefix\endcsname\relax\def\urlprefix{URL }\fi
\providecommand{\bibinfo}[2]{#2}
\providecommand{\eprint}[2][]{\url{#2}}

\bibitem[{\citenamefont{Ritchie}(1957)}]{Ritchie_1957}
\bibinfo{author}{\bibfnamefont{R.~H.} \bibnamefont{Ritchie}},
  \bibinfo{journal}{Phys. Rev.} \textbf{\bibinfo{volume}{106}},
  \bibinfo{pages}{874} (\bibinfo{year}{1957}).

\bibitem[{\citenamefont{Powell and
  Swan}(1959{\natexlab{a}})}]{Powell_Swan_1959_1}
\bibinfo{author}{\bibfnamefont{C.~J.} \bibnamefont{Powell}} \bibnamefont{and}
  \bibinfo{author}{\bibfnamefont{J.~B.} \bibnamefont{Swan}},
  \bibinfo{journal}{Phys. Rev.} \textbf{\bibinfo{volume}{115}},
  \bibinfo{pages}{869} (\bibinfo{year}{1959}{\natexlab{a}}).

\bibitem[{\citenamefont{Powell and
  Swan}(1959{\natexlab{b}})}]{Powell_Swan_1959_2}
\bibinfo{author}{\bibfnamefont{C.~J.} \bibnamefont{Powell}} \bibnamefont{and}
  \bibinfo{author}{\bibfnamefont{J.~B.} \bibnamefont{Swan}},
  \bibinfo{journal}{Phys. Rev.} \textbf{\bibinfo{volume}{116}},
  \bibinfo{pages}{81} (\bibinfo{year}{1959}{\natexlab{b}}).

\bibitem[{\citenamefont{Otto}(1968)}]{Otto_1968}
\bibinfo{author}{\bibfnamefont{A.}~\bibnamefont{Otto}}, \bibinfo{journal}{Z.
  Phys.} \textbf{\bibinfo{volume}{216}}, \bibinfo{pages}{398}
  (\bibinfo{year}{1968}).

\bibitem[{\citenamefont{Kretschmann and Raether}(1968)}]{Kretschmann_1968}
\bibinfo{author}{\bibfnamefont{E.}~\bibnamefont{Kretschmann}} \bibnamefont{and}
  \bibinfo{author}{\bibfnamefont{H.}~\bibnamefont{Raether}},
  \bibinfo{journal}{Z. Naturf. A} \textbf{\bibinfo{volume}{23}},
  \bibinfo{pages}{2135} (\bibinfo{year}{1968}).

\bibitem[{\citenamefont{Trivelpiece and Gould}(1959)}]{Trivelpiece_Gould_1959}
\bibinfo{author}{\bibfnamefont{A.~W.} \bibnamefont{Trivelpiece}}
  \bibnamefont{and} \bibinfo{author}{\bibfnamefont{R.~W.} \bibnamefont{Gould}},
  \bibinfo{journal}{J. Appl. Phys.} \textbf{\bibinfo{volume}{30}},
  \bibinfo{pages}{1784} (\bibinfo{year}{1959}).

\bibitem[{\citenamefont{Guernsey}(1969)}]{Guernsey_1969}
\bibinfo{author}{\bibfnamefont{R.~L.} \bibnamefont{Guernsey}},
  \bibinfo{journal}{Phys. Fluids} \textbf{\bibinfo{volume}{12}},
  \bibinfo{pages}{1852} (\bibinfo{year}{1969}).

\bibitem[{\citenamefont{Kondratenko}(1985)}]{Kondratenko_book}
\bibinfo{author}{\bibfnamefont{A.~N.} \bibnamefont{Kondratenko}},
  \emph{\bibinfo{title}{Surface and Volume Waves in Bounded Plasmas}}
  (\bibinfo{publisher}{Energoatomizdat}, \bibinfo{year}{1985}).

\bibitem[{\citenamefont{Vladimirov et~al.}(1994)\citenamefont{Vladimirov, Yu,
  and Tsytovich}}]{Vladimirov_progress_1994}
\bibinfo{author}{\bibfnamefont{S.~V.} \bibnamefont{Vladimirov}},
  \bibinfo{author}{\bibfnamefont{M.~Y.} \bibnamefont{Yu}}, \bibnamefont{and}
  \bibinfo{author}{\bibfnamefont{V.~N.} \bibnamefont{Tsytovich}},
  \bibinfo{journal}{Phys. Rep.} \textbf{\bibinfo{volume}{241}},
  \bibinfo{pages}{1} (\bibinfo{year}{1994}).

\bibitem[{\citenamefont{Denysenko et~al.}(2002)\citenamefont{Denysenko, Gapon,
  Azarenkov, Ostrikov, and Yu}}]{Denysenko_etal_2002}
\bibinfo{author}{\bibfnamefont{I.~B.} \bibnamefont{Denysenko}},
  \bibinfo{author}{\bibfnamefont{A.~V.} \bibnamefont{Gapon}},
  \bibinfo{author}{\bibfnamefont{N.~A.} \bibnamefont{Azarenkov}},
  \bibinfo{author}{\bibfnamefont{K.~N.} \bibnamefont{Ostrikov}},
  \bibnamefont{and} \bibinfo{author}{\bibfnamefont{M.~Y.} \bibnamefont{Yu}},
  \bibinfo{journal}{Phys. Rev. E} \textbf{\bibinfo{volume}{65}},
  \bibinfo{pages}{046419} (\bibinfo{year}{2002}).

\bibitem[{\citenamefont{Shokri}(2002)}]{Shokri_2002}
\bibinfo{author}{\bibfnamefont{B.}~\bibnamefont{Shokri}},
  \bibinfo{journal}{Phys. Plasmas} \textbf{\bibinfo{volume}{7}},
  \bibinfo{pages}{701} (\bibinfo{year}{2002}).

\bibitem[{\citenamefont{Pitarke et~al.}(2007)\citenamefont{Pitarke, Silkin,
  Chulkov, and Echenique}}]{Pitarke_etal_2007}
\bibinfo{author}{\bibfnamefont{J.~M.} \bibnamefont{Pitarke}},
  \bibinfo{author}{\bibfnamefont{V.~M.} \bibnamefont{Silkin}},
  \bibinfo{author}{\bibfnamefont{E.~V.} \bibnamefont{Chulkov}},
  \bibnamefont{and} \bibinfo{author}{\bibfnamefont{P.~M.}
  \bibnamefont{Echenique}}, \bibinfo{journal}{Rep. Prog. Phys.}
  \textbf{\bibinfo{volume}{70}}, \bibinfo{pages}{1} (\bibinfo{year}{2007}).

\bibitem[{\citenamefont{Ozbay}(2006)}]{Ozbay_2006}
\bibinfo{author}{\bibfnamefont{E.}~\bibnamefont{Ozbay}},
  \bibinfo{journal}{Science} \textbf{\bibinfo{volume}{311}},
  \bibinfo{pages}{189} (\bibinfo{year}{2006}).

\bibitem[{\citenamefont{Barnes et~al.}(2003)\citenamefont{Barnes, Dereux, and
  Ebbesen}}]{Barnes_etal_2003}
\bibinfo{author}{\bibfnamefont{W.~L.} \bibnamefont{Barnes}},
  \bibinfo{author}{\bibfnamefont{A.}~\bibnamefont{Dereux}}, \bibnamefont{and}
  \bibinfo{author}{\bibfnamefont{T.~W.} \bibnamefont{Ebbesen}},
  \bibinfo{journal}{Nature} \textbf{\bibinfo{volume}{424}},
  \bibinfo{pages}{824} (\bibinfo{year}{2003}).

\bibitem[{\citenamefont{Nomura et~al.}(2005)\citenamefont{Nomura, Ohtsu, and
  Yatsui}}]{Nomura_etal_2005}
\bibinfo{author}{\bibfnamefont{W.}~\bibnamefont{Nomura}},
  \bibinfo{author}{\bibfnamefont{M.}~\bibnamefont{Ohtsu}}, \bibnamefont{and}
  \bibinfo{author}{\bibfnamefont{T.}~\bibnamefont{Yatsui}},
  \bibinfo{journal}{Appl. Phys. Lett.} \textbf{\bibinfo{volume}{86}},
  \bibinfo{pages}{181108} (\bibinfo{year}{2005}).

\bibitem[{\citenamefont{Brongersma and
  Shalaev}(2010)}]{Plasmonics_Science_2010}
\bibinfo{author}{\bibfnamefont{M.~L.} \bibnamefont{Brongersma}}
  \bibnamefont{and} \bibinfo{author}{\bibfnamefont{V.~M.}
  \bibnamefont{Shalaev}}, \bibinfo{journal}{Science}
  \textbf{\bibinfo{volume}{328}}, \bibinfo{pages}{440} (\bibinfo{year}{2010}).

\bibitem[{\citenamefont{Bergman and Stockman}(2003)}]{SPASER_PRL_2003}
\bibinfo{author}{\bibfnamefont{D.~J.} \bibnamefont{Bergman}} \bibnamefont{and}
  \bibinfo{author}{\bibfnamefont{M.~I.} \bibnamefont{Stockman}},
  \bibinfo{journal}{Phys. Rev. Lett.} \textbf{\bibinfo{volume}{90}},
  \bibinfo{pages}{027402} (\bibinfo{year}{2003}).

\bibitem[{\citenamefont{Zheludev et~al.}(2008)\citenamefont{Zheludev,
  Prosvirnin, Parasimakis, and Fedotov}}]{Zheludev_etal_2008}
\bibinfo{author}{\bibfnamefont{N.~I.} \bibnamefont{Zheludev}},
  \bibinfo{author}{\bibfnamefont{S.~L.} \bibnamefont{Prosvirnin}},
  \bibinfo{author}{\bibfnamefont{N.}~\bibnamefont{Parasimakis}},
  \bibnamefont{and} \bibinfo{author}{\bibfnamefont{V.~A.}
  \bibnamefont{Fedotov}}, \bibinfo{journal}{Nature Photon.}
  \textbf{\bibinfo{volume}{2}}, \bibinfo{pages}{351} (\bibinfo{year}{2008}).

\bibitem[{\citenamefont{Noginov et~al.}(2009)\citenamefont{Noginov, Zhu,
  Belgrave, Bakker, Shalaev, Narimanov, Stout, Herz, Suteewong, and
  Wiesner}}]{Noginov_etal_2009}
\bibinfo{author}{\bibfnamefont{M.~A.} \bibnamefont{Noginov}},
  \bibinfo{author}{\bibfnamefont{G.}~\bibnamefont{Zhu}},
  \bibinfo{author}{\bibfnamefont{A.~M.} \bibnamefont{Belgrave}},
  \bibinfo{author}{\bibfnamefont{R.}~\bibnamefont{Bakker}},
  \bibinfo{author}{\bibfnamefont{V.~M.} \bibnamefont{Shalaev}},
  \bibinfo{author}{\bibfnamefont{E.~E.} \bibnamefont{Narimanov}},
  \bibinfo{author}{\bibfnamefont{S.}~\bibnamefont{Stout}},
  \bibinfo{author}{\bibfnamefont{E.}~\bibnamefont{Herz}},
  \bibinfo{author}{\bibfnamefont{T.}~\bibnamefont{Suteewong}},
  \bibnamefont{and} \bibinfo{author}{\bibfnamefont{U.}~\bibnamefont{Wiesner}},
  \bibinfo{journal}{Nature} \textbf{\bibinfo{volume}{460}},
  \bibinfo{pages}{1110} (\bibinfo{year}{2009}).

\bibitem[{\citenamefont{Garcia-Vidal and Moreno}(2009)}]{Lasers_go_nano}
\bibinfo{author}{\bibfnamefont{F.~J.} \bibnamefont{Garcia-Vidal}}
  \bibnamefont{and} \bibinfo{author}{\bibfnamefont{E.}~\bibnamefont{Moreno}},
  \bibinfo{journal}{Nature} \textbf{\bibinfo{volume}{461}},
  \bibinfo{pages}{604} (\bibinfo{year}{2009}).

\bibitem[{\citenamefont{Vladimirov}(1994)}]{Vladimirov_Kohn_1994}
\bibinfo{author}{\bibfnamefont{S.~V.} \bibnamefont{Vladimirov}},
  \bibinfo{journal}{Phys. Scr.} \textbf{\bibinfo{volume}{49}},
  \bibinfo{pages}{625} (\bibinfo{year}{1994}).

\bibitem[{\citenamefont{Marklund et~al.}(2008)\citenamefont{Marklund, Brodin,
  Stenflo, and Liu}}]{Marklund_etal_new_quantum_limits}
\bibinfo{author}{\bibfnamefont{M.}~\bibnamefont{Marklund}},
  \bibinfo{author}{\bibfnamefont{G.}~\bibnamefont{Brodin}},
  \bibinfo{author}{\bibfnamefont{L.}~\bibnamefont{Stenflo}}, \bibnamefont{and}
  \bibinfo{author}{\bibfnamefont{C.~S.} \bibnamefont{Liu}},
  \bibinfo{journal}{Europhys. Lett.} \textbf{\bibinfo{volume}{84}},
  \bibinfo{pages}{17006} (\bibinfo{year}{2008}).

\bibitem[{\citenamefont{Tyshetskiy et~al.}(2012)\citenamefont{Tyshetskiy,
  Williamson, Kompaneets, and Vladimirov}}]{Tysh_etal_ESSW_2012}
\bibinfo{author}{\bibfnamefont{Y.}~\bibnamefont{Tyshetskiy}},
  \bibinfo{author}{\bibfnamefont{D.~J.} \bibnamefont{Williamson}},
  \bibinfo{author}{\bibfnamefont{R.}~\bibnamefont{Kompaneets}},
  \bibnamefont{and} \bibinfo{author}{\bibfnamefont{S.~V.}
  \bibnamefont{Vladimirov}}, \bibinfo{journal}{Phys. Plasmas}
  \textbf{\bibinfo{volume}{19}}, \bibinfo{pages}{032102}
  (\bibinfo{year}{2012}).

\bibitem[{\citenamefont{Vladimirov and Tyshetskiy}(2011)}]{Vlad_Tysh_UFN_2011}
\bibinfo{author}{\bibfnamefont{S.~V.} \bibnamefont{Vladimirov}}
  \bibnamefont{and} \bibinfo{author}{\bibfnamefont{Y.~O.}
  \bibnamefont{Tyshetskiy}}, \bibinfo{journal}{Phys. Usp., in press}
  (\bibinfo{year}{2011}).

\bibitem[{\citenamefont{H.~W.~Wyld and Pines}(1962)}]{Wyld_Pines_1962}
\bibinfo{author}{\bibfnamefont{J.}~\bibnamefont{H.~W.~Wyld}} \bibnamefont{and}
  \bibinfo{author}{\bibfnamefont{D.}~\bibnamefont{Pines}},
  \bibinfo{journal}{Phys. Rev.} \textbf{\bibinfo{volume}{127}},
  \bibinfo{pages}{1851} (\bibinfo{year}{1962}).

\bibitem[{\citenamefont{Kremp et~al.}(2005)\citenamefont{Kremp, Schlanges, and
  Kraeft}}]{Kremp_book}
\bibinfo{author}{\bibfnamefont{D.}~\bibnamefont{Kremp}},
  \bibinfo{author}{\bibfnamefont{M.}~\bibnamefont{Schlanges}},
  \bibnamefont{and} \bibinfo{author}{\bibfnamefont{W.-D.}
  \bibnamefont{Kraeft}}, \emph{\bibinfo{title}{Quantum Statistics of Nonideal
  Plasmas}} (\bibinfo{publisher}{Springer}, \bibinfo{year}{2005}).

\bibitem[{\citenamefont{Ziman}(1961)}]{Ziman_1961}
\bibinfo{author}{\bibfnamefont{J.~M.} \bibnamefont{Ziman}},
  \bibinfo{journal}{Philos. Mag.} \textbf{\bibinfo{volume}{6}},
  \bibinfo{pages}{1013} (\bibinfo{year}{1961}).

\bibitem[{\citenamefont{Alekseev et~al.}(1972)\citenamefont{Alekseev, Andreev,
  and Prokhorenko}}]{Alekseev_UFN}
\bibinfo{author}{\bibfnamefont{V.~A.} \bibnamefont{Alekseev}},
  \bibinfo{author}{\bibfnamefont{A.~A.} \bibnamefont{Andreev}},
  \bibnamefont{and} \bibinfo{author}{\bibfnamefont{V.~Y.}
  \bibnamefont{Prokhorenko}}, \bibinfo{journal}{Sov. Phys. Usp.}
  \textbf{\bibinfo{volume}{15}}, \bibinfo{pages}{139} (\bibinfo{year}{1972}).

\bibitem[{\citenamefont{Alexandrov et~al.}(1984)\citenamefont{Alexandrov,
  Bogdankevich, and Rukhadze}}]{ABR_book}
\bibinfo{author}{\bibfnamefont{A.~F.} \bibnamefont{Alexandrov}},
  \bibinfo{author}{\bibfnamefont{L.~S.} \bibnamefont{Bogdankevich}},
  \bibnamefont{and} \bibinfo{author}{\bibfnamefont{A.~A.}
  \bibnamefont{Rukhadze}}, \emph{\bibinfo{title}{Principles of Plasma
  Electrodynamics}} (\bibinfo{publisher}{Springer-Verlag},
  \bibinfo{year}{1984}).

\bibitem[{\citenamefont{Lide}(2008)}]{Phys_Chem_book}
\bibinfo{author}{\bibfnamefont{D.~R.} \bibnamefont{Lide}},
  \emph{\bibinfo{title}{CRC Handbook of Chemistry and Physics}}
  (\bibinfo{publisher}{CRC Press}, \bibinfo{year}{2008}).

\bibitem[{\citenamefont{Ordal et~al.}(1985)\citenamefont{Ordal, Bell,
  R.~W.~Alexander, Long, and Querry}}]{Ordal_etal_1985}
\bibinfo{author}{\bibfnamefont{M.~A.} \bibnamefont{Ordal}},
  \bibinfo{author}{\bibfnamefont{R.~J.} \bibnamefont{Bell}},
  \bibinfo{author}{\bibfnamefont{J.}~\bibnamefont{R.~W.~Alexander}},
  \bibinfo{author}{\bibfnamefont{L.~L.} \bibnamefont{Long}}, \bibnamefont{and}
  \bibinfo{author}{\bibfnamefont{M.~R.} \bibnamefont{Querry}},
  \bibinfo{journal}{Appl. Opt.} \textbf{\bibinfo{volume}{24}},
  \bibinfo{pages}{4493} (\bibinfo{year}{1985}).

\end{thebibliography}

\end{document}